\documentclass[aps,prl,amsmath,amssymb,nofootinbib,twocolumn]{revtex4}
\usepackage{amssymb}
\usepackage{color}
\usepackage{graphicx}
\usepackage{subfigure}
\usepackage{epsfig,amssymb,amsmath}
\usepackage[hypertex,linkcolor=red]{hyperref}

\usepackage{color} 

\def\>{\rangle}
\def\<{\langle}

\def\labell#1{\label{#1}}
\def\section#1{{\par\em #1:--- }}

\begin{document}


\title{A simple proof of Bell's inequality} \author{Lorenzo Maccone}
\affiliation{ \vbox{Dip.~Fisica, Univ.~of Pavia, via Bassi 6, I-27100
    Pavia, Italy} }

\begin{abstract}
  Bell's theorem is a fundamental result in quantum mechanics: it
  discriminates between quantum mechanics and all theories where
  probabilities in measurement results arise from the ignorance of
  pre-existing local properties. We give an extremely simple proof of
  Bell's inequality: a single figure suffices. This simplicity may be
  useful in the unending debate of what exactly the Bell inequality
  means, since the hypothesis at the basis of the proof become
  extremely transparent. It is also a useful didactic tool, as the
  Bell inequality can be explained in a single intuitive lecture.
\end{abstract}
\maketitle

\section{Introduction}
Einstein had a dream. He believed quantum mechanics was an incomplete
description of reality \cite{epr} and that its completion might
explain the troublesome fundamental probabilities of quantum mechanics
as emerging from some hidden degrees of freedom: probabilities would
arise because of our ignorance of these ``hidden variables''. His
dream was that probabilities in quantum mechanics might turn out to
have the same meaning as probabilities in classical thermodynamics,
where they refer to our ignorance of the microscopic degrees of
freedom (e.g.~the position and velocity of each gas molecule): he
wrote, ``the statistical quantum theory would, within the framework of
future physics, take an approximately analogous position to the
statistical mechanics within the framework of classical
mechanics''~\cite{schilpp}.

A decade after Einstein's death, John Bell shattered this dream
\cite{bell,speakable,merminrmp}: any completion of quantum mechanics with
hidden variables would be incompatible with relativistic causality!
The essence of Bell's theorem is that quantum mechanical probabilities
cannot arise from the ignorance of {\em local} pre-existing variables.
In other words, if we want to assign pre-existing (but hidden)
properties to explain probabilities in quantum measurements, these
properties must be non-local: an agent with access to the non-local
variables could transmit information instantly to a distant location,
thus violating relativistic causality and awakening the nastiest
temporal paradoxes \cite{werner}.

[It is important to emphasize that we use ``local'' here in Einstein's
connotation: locality implies superluminal communication is
impossible. In contrast, often quantum mechanics is deemed
``non-local'' in the sense that correlations among properties can
propagate instantly, thanks to entanglement \cite{epr}. This `quantum
non-locality' cannot be used to transfer information instantly as
correlations cannot be used to that aim. In the remainder of the paper
we will only use the former meaning of locality (Einstein
non-locality) and we warn the reader not to confuse it with the latter
(quantum non-locality).]

Modern formulations of quantum mechanics must incorporate Bell's
result at their core: either they refuse the idea that measurements
uncover pre-existing properties, or they must make use of non-local
properties. In the latter case, they must also introduce some
censorship mechanism to prevent the use of hidden variables to
transmit information. An example of the first formulation is the
conventional Copenhagen interpretation of quantum mechanics, which
(thanks to complementarity) states that the properties arise from the
interaction between the quantum system and the measurement apparatus,
they are not pre-existing: ``unperformed experiments have no results''
\cite{peres}. An example of the second formulation is the ``de
Broglie-Bohm interpretation'' of quantum mechanics that assumes that
particle trajectories are hidden variables (they exist independently
of position measurements).

Bell's result is at the core of modern quantum mechanics, as it
elucidates the theory's precarious co-existence with relativistic
causality. It has spawned an impressive amount of research. However,
it is often ignored in basic quantum mechanics courses since
traditional proofs of Bell's theorem are rather cumbersome and often
overburdened by philosophical considerations. Here we give an
extremely simple graphical proof of Mermin's version
\cite{mermin,preskill} of Bell's theorem. The simplicity of the proof
is key to clarifying all the theorem's assumptions, the identification
of which generated a large debate in the literature (e.g.~see
\cite{auletta}). Here we focus on simplifying of the proof. We refer
the reader that wants to gain an intuition of the quantum part to
Refs.~\cite{kwiat,wiseman}, and to \cite{macdon} for a proof without
probabilities.

\section{Bell's theorem}
Let us define ``local'' a theory where the outcomes of an experiment
on a system are independent of the actions performed on a different
system which has no causal connection with the first. [As stated
previously, this refers to locality in Einstein's connotation of the
word: the outcomes of the experiment cannot be used to receive
information from whoever acts on the second system, if it has no
causal connection to the first.] For example, the temperature of my
room is independent on whether you choose to wear a purple tie today.
Einstein's relativity provides a stringent condition for causal
connections: if two events are outside their respective light cones,
there cannot be any causal connection among them.

Let us define ``counterfactual-definite'' \cite{perbell,stapp} a
theory whose experiments uncover properties that are pre-existing. In
other words, in a counterfactual-definite theory it is meaningful to
assign a property to a system (e.g.~the position of an electron)
independently of whether the measurement of such property is carried
out. [Sometime this counterfactual definiteness property is also
called ``realism'', but it is best to avoid such philosophically laden
term to avoid misconceptions.]

Bell's theorem can be phrased as ``quantum mechanics cannot be both
local and counterfactual-definite''. A logically equivalent way of stating it
is ``quantum mechanics is either non-local or non counterfactual-definite''.

To prove this theorem, Bell provided an inequality (referring to
correlations of measurement results) that is satisfied by all theories
that are both local {\em and} counterfactual-definite. He then showed
that quantum mechanics violates this inequality, and hence cannot be
local and counterfactual-definite. 

It is important to note that the Bell inequality can be derived also
using weaker hypotheses than ``Einstein locality'' and
``counterfactual definiteness'': such a proof is presented in Appendix
A (where Einstein locality is relaxed to ``Bell locality'' and
counterfactual definiteness is relaxed to ``hidden variable models'').
However, from a physical point of view, the big impact of Bell's
theorem is to prove the incompatibility of quantum mechanics with
local counterfactual-definite properties, and we will stick to these
hypotheses in the main text (see also Appendix B for a schematic
formalization of all these results).

A couple of additional hypothesis at the basis of Bell's theorem are
often left implicit: (1) our choice of which experiment to perform
must be independent of the properties of the object to be measured
(technically, ``freedom of choice'' or ``no super-determinism''
\cite{speakable}): e.g., if we decided to measure the color of red
objects only, we would falsely conclude that all objects are red; (2)
future outcomes of the experiment must not influence which apparatus
settings were previously chosen \cite{botero} (whereas clearly the
apparatus settings will influence the outcomes): a trivial causality
requirement (technically, ``measurement independence'').  These two
hypothesis are usually left implicit because science would be
impossible without them.

All experiments performed to date (e.g.~\cite{grangier,kwiatex,zeil})
have shown that Bell inequalities are violated, suggesting that our
world cannot be both local and counterfactual-definite. However, it
should be noted that no experiment up to now has been able to test
Bell inequalities rigorously, because additional assumptions are
required to take care of experimental imperfections. These assumptions
are all quite reasonable, so that only conspiratorial alternatives to
quantum mechanics have yet to be ruled out (where experimental
imperfections are fine-tuned to the properties of the objects
\cite{pearle}, namely they violate the ``freedom of choice''). In the
next couple of years the definitive Bell inequality experiment will be
performed: many research groups worldwide are actively pursuing it.

\begin{figure}[t]
\begin{center}
  \epsfxsize=.9\hsize\leavevmode\epsffile{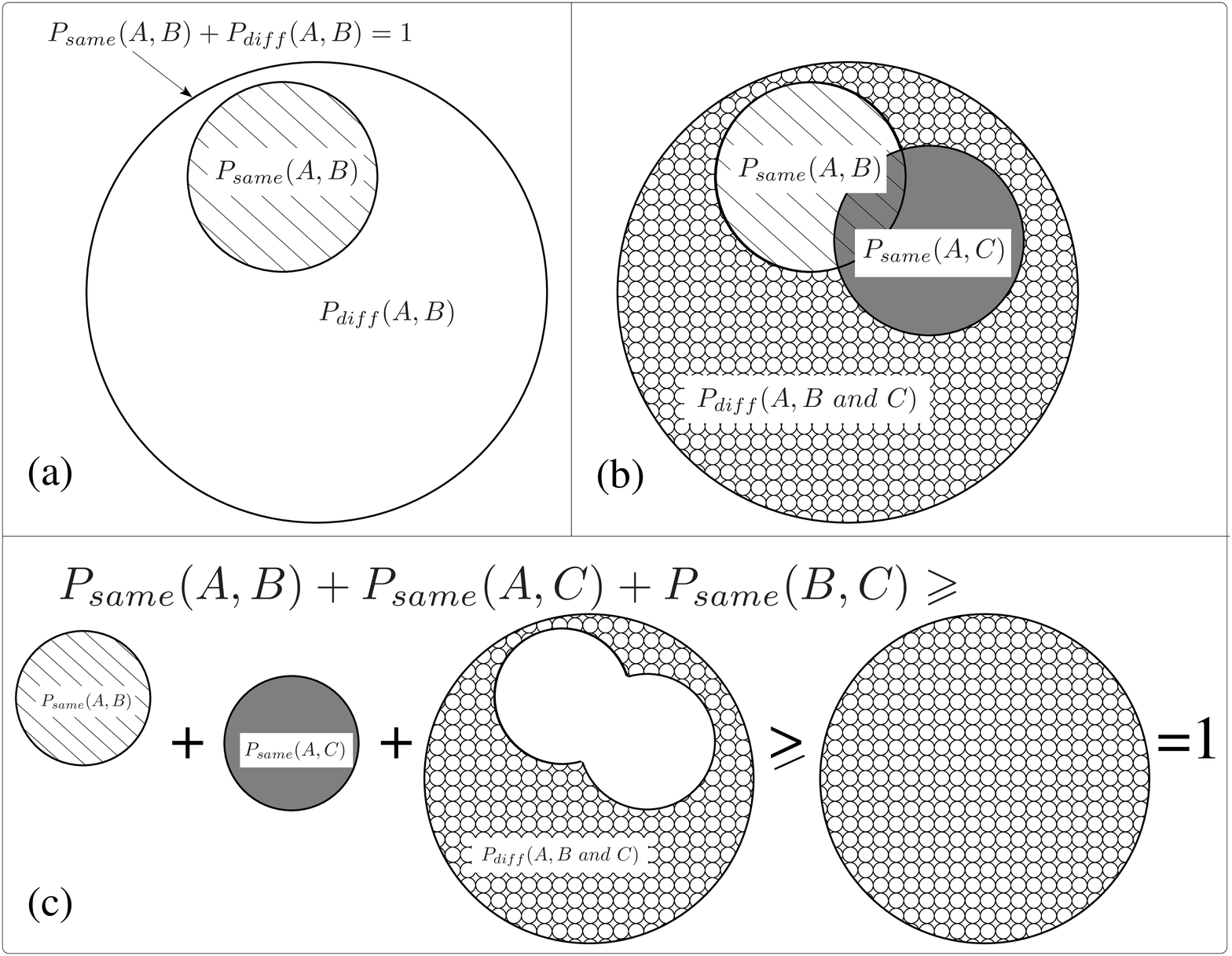}
\end{center}
\vspace{-.5cm}
\caption{ Proof of Bell inequality \eqref{bellineq} using areas to
  represent probabilities. (a)~The dashed area represents the
  probability that property $A$ of the first object and $B$ of the
  second are equal (both 1 or both 0): $P_{same}(A,B)$. The white area
  represents the probability that they are different: $P_{diff}(A,B)$.
  The whole circle has area $1=P_{same}(A,B)+P_{diff}(A,B)$. (b)~The
  gray area represents the probability that $A$ and $C$ are equal, and
  the non-gray area represents the probability that $A$ and $C$ are
  different. If $A$ of the first object is different from both $B$ and
  $C$ of the second (dotted area), then $B$ and $C$ of the second
  object must be the same. Hence, the probability that $B$ and $C$ are
  the same must be larger than (or equal to) the dotted area: since
  $B$ is the same for the two objects, $P_{same}(B,C)$ must be larger
  than (or equal to) the dotted area.  (c)~The quantity
  $P_{same}(A,B)+P_{same}(A,C)+P_{same}(B,C)$ is hence larger than (or
  equal to) the sum of the dashed $+$ gray $+$ dotted areas, which is
  in turn larger than (or equal to) the full circle of area 1: this
  proves the Bell inequality \eqref{bellineq}.  The reasoning fails if
  we do not employ counterfactual-definite properties, for example if
  complementarity prevents us from assigning values to both properties
  $B$ and $C$ of the second object.  It also fails if we employ
  non-local properties, for example if a measurement of $B$ on an
  object to find its value changes the value of $A$ of the other
  object.  }
\labell{f:p}\end{figure}

\section{Proof of Bell's theorem} We use the Bell inequality proposed
by Preskill \cite{preskill}, following Mermin's suggestion
\cite{mermin}. Suppose we have two identical objects, namely they have
the same properties. Suppose also that these properties are
predetermined (counterfactual definiteness) and not generated by their
measurement, and that the determination of the properties of one
object will not influence any property of the other object (locality).

We will only need three properties $A$, $B$, and $C$ that can each
take two values: ``0'' and ``1''. For example, if the objects are
coins, then $A=0$ might mean that the coin is gold and $A=1$ that the
coin is copper (property $A$, material), $B=0$ means the coin is shiny
and $B=1$ it is dull (property $B$, texture), and $C=0$ means the coin
is large and $C=1$ it is small (property $C$, size).  

Suppose I do not know the properties because the two coins are a gift
in two wrapped boxes: I only know the gift is two identical coins, but
I do not know whether they are two gold, shiny, small coins
$(A=0,B=0,C=1)$ or two copper, shiny, large coins $(1,0,0)$ or two
gold, dull, large coins $(1,1,0)$, etc. I do know that the properties
``exist'' (namely, they are counterfactual-definite and predetermined even if I
cannot see them directly) and they are local (namely, acting on one
box will not change any property of the coin in the other box: the
properties refer separately to each coin). These are quite reasonable
assumptions for two coins!  My ignorance of the properties is
expressed through probabilities that represent either my expectation
of finding a property (Bayesian view), or the result of performing
many repeated experiments with boxes and coins and averaging over some
possibly hidden variable, typically indicated with the letter
$\lambda$ \cite{speakable}, that determines the property (frequentist
view) \cite{peres}.  For example, I might say the gift bearer will
give me two gold coins with a $20\%$ probability (he is stingy, but
not always).

Bell's inequality refers to the correlation among measurement outcomes
of the properties: call $P_{same}(A,B)$ the probability that the
properties $A$ of the first object and $B$ of the second are the same:
$A$ and $B$ are both 0 (the first coin is gold and the second is
shiny) or they are both 1 (the first is copper and the second is
dull). For example, $P_{same}(A,B)=1/2$ tells me that with $50\%$
chance $A=B$ (namely they are both 0 or both 1). Since the two coins
have equal counterfactual-definite properties, this also implies that
with $50\%$ chance I get two gold shiny coins or two copper dull
coins. Note that the fact that the two coins have the same properties
means that $P_{same}(A,A)=P_{same}(B,B)=P_{same}(C,C)=1$: if one is
made of gold, also the other one will be, or if one is made of copper,
also the other one will be, etc.

{\em Bell's inequality \cite{preskill}:} Under the conditions that
three arbitrary two-valued properties $A$, $B$, $C$ satisfy
counterfactual definiteness and locality, and that $P_{same}(X,X)=1$
for $X=A,B,C$ (i.e.~the two objects have same properties), the
following inequality among correlations holds,
\begin{eqnarray}
  P_{same}(A,B)+
  P_{same}(A,C)+
  P_{same}(B,C)\geqslant 1
\labell{bellineq}\;,
\end{eqnarray}
namely, a Bell inequality. The proof of such inequality is given
graphically in Fig.~\ref{f:p}. The inequality basically says that the
sum of the probabilities that the two properties are the same if I
consider respectively $A$ and $B$, $A$ and $C$, and $B$ and $C$ must
be larger than one. This is intuitively clear: since the two coins
have the same properties, the sum of the probabilities that the coins
are gold and shiny, copper and dull, gold and large, copper and small,
shiny and small, dull and large is greater than one: all the
combinations have been counted, possibly more than once. [In
Fig.~\ref{f:a} the events to which the probabilities represented by
the Venn diagrams of Fig.~\ref{f:p} refer are made explicit.]

\begin{figure}[hbt]
\begin{center}
  \epsfxsize=.7\hsize\leavevmode\epsffile{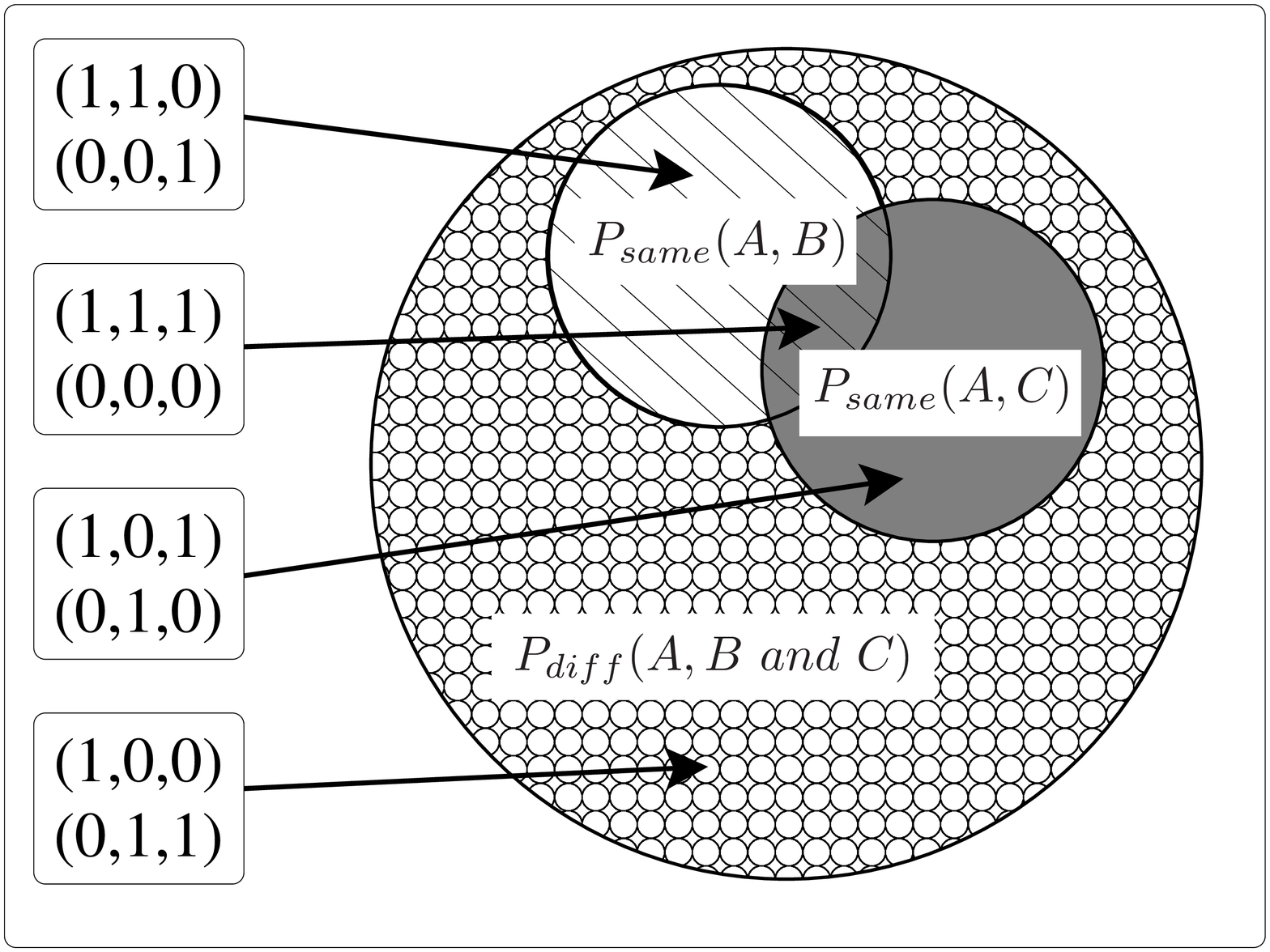}
\end{center}
\vspace{-.5cm}
\caption{ Explicit depiction of the properties whose probabilities are
  represented by the areas of the Venn diagrams in Fig.~\ref{f:p}. The
  properties are represented by a triplet of numbers $(A,B,C)$ that
  indicate the (counterfactual-definite, local) values of the properties $A$,
  $B$, and $C$ for both objects.  Note that in the dotted area $A$
  must be different from both $B$ and $C$, so that $B$ and $C$ must be
  equal there ($B$ and $C$ are equal also in the intersection between
  the two smaller sets, but that is irrelevant to the proof).}
\labell{f:a}\end{figure}

This is true, of course, only if the two objects have same
counterfactual-definite properties and the measurement of one does not
affect the outcome of the other. If we lack counterfactual-definite
properties, we cannot infer that the first coin is shiny only because
we measured the second to be shiny, even if we know that the two coins
have the same properties: without counterfactual definiteness, we
cannot even speak of the first coin's texture unless we measure it.
Moreover, if a measurement of the second coin's texture can change the
one of the first coin (non-locality) again we cannot infer the first
coin's texture from a measurement of the second: even if we know that
the initial texture of the coins was the same, the measurement on the
second may change such property of the first. Both the
``counterfactual definiteness'' and the ``Einstein locality''
hypotheses we used here can be relaxed somewhat, as shown in Appendix
A (suggested only to more advanced readers).

To prove Bell's theorem, we now provide a quantum system that violates
the above inequality. Consider two two-level systems (qubits) in the
joint entangled state $|\Phi^+\>=(|00\>+|11\>)/\sqrt{2}$, and consider
the two-valued properties $A$, $B$, and $C$ obtained by projecting the
qubit on the states 
\begin{eqnarray}
&&A:
\left\{\begin{array}{lll}
|a_0\rangle&\equiv&|0\> \cr
|a_1\rangle&\equiv&|1\>\;,
\end{array}\right.
\qquad B:
\left\{\begin{array}{lll}
|b_0\rangle&\equiv& \frac 12 |0\rangle+\frac{\sqrt{3}}2|1\rangle\cr
|b_1\rangle&\equiv& \frac {\sqrt{3}}2 |0\rangle
-\frac 12|1\rangle\;,
\end{array}\right.\nonumber\\&&
C:\left\{\begin{array}{lll}
|c_0\rangle&\equiv& \frac 12 |0\rangle-\frac{\sqrt{3}}2|1\rangle\cr
|c_1\rangle&\equiv& \frac {\sqrt{3}}2 |0\rangle+\frac 12|1\rangle\;,
\end{array}\right.
\labell{prop}\;
\end{eqnarray}
where it is easy to check that $|b_1\>$ is orthogonal to $|b_0\>$ and
$|c_1\>$ is orthogonal to $|c_0\>$. It is also easy to check that 
\begin{eqnarray}
&&|\Phi^+\>=\labell{pa}\\\nonumber&&
\frac{|a_0a_0\>+|a_1a_1\>}{\sqrt{2}}=\frac{|b_0b_0\>+|b_1b_1\>}{\sqrt{2}}=
\frac{|c_0c_0\>+|c_1c_1\>}{\sqrt{2}}
\;,
\end{eqnarray}
so that the two qubits have the same properties, namely
$P_{same}(A,A)=P_{same}(B,B)=P_{same}(C,C)=1$: the measurement of the
same property on both qubits always yields the same outcome, both 0 or
both 1. 

We are now ready to calculate the quantity on the left of Bell's
inequality \eqref{bellineq}. Just write the state $|\Phi^+\>$ in terms
of the eigenstates of the properties $A$, $B$ and $C$. E.g., it is
easy to find the value of $P_{same}(A,B)$ if we write
\begin{eqnarray}
|\Phi^+\rangle=\frac
{|a_0\rangle(|b_0\rangle+\sqrt{3}|b_1\rangle)+
|a_1\rangle(|\sqrt{3}|b_0\rangle-|b_1\rangle)}{2\sqrt{2}}
\nonumber\;.
\end{eqnarray}
In fact, the probability of obtaining zero for both properties is the
square modulus of the coefficient of $|a_0\>|b_0\>$, namely
$|1/2\sqrt{2}|^2=1/8$, while the probability of obtaining one for both
is the square modulus of the coefficient of $|a_1\>|b_1\>$, again
$1/8$. Hence, $P_{same}(A,B)=1/8+1/8=1/4$. Analogously, we find that
$P_{same}(A,C)=1/4$ and that $P_{same}(B,C)=1/4$ by expressing the
state as  
\begin{eqnarray}
|\Phi^+\rangle&=&\tfrac
{|a_0\rangle(|c_0\rangle+\sqrt{3}|c_1\rangle)-
|a_1\rangle(|\sqrt{3}|c_0\rangle-|c_1\rangle)}{2\sqrt{2}}
\nonumber
\\|\Phi^+\rangle&=&\tfrac{
(|b_0\rangle+\sqrt{3}|b_1\rangle)
(|c_0\rangle+\sqrt{3}|c_1\rangle)-(\sqrt{3}|b_0\rangle-|b_1\rangle)
(\sqrt{3}|c_0\rangle-|c_1\rangle)}
{4\sqrt{2}}
\nonumber.
\end{eqnarray}
Summarizing, we have found
\begin{eqnarray}
P_{same}(A,B)+P_{same}(A,C)+P_{same}(B,C)=\tfrac34<1
\labell{violaz}\;,
\end{eqnarray}
which violates Bell's inequality \eqref{bellineq}.

This proves Bell's theorem: all theories that are both local and
counterfactual-definite must satisfy inequality \eqref{bellineq} which
is violated by quantum mechanics.  Then, quantum mechanics cannot be a
local counterfactual-definite theory: it must either be
non-counterfactual-definite (as in the Copenhagen interpretation) or
non-local (as in the de Broglie-Bohm interpretation) \cite{bellc}.

\section{Acknowledgments} I acknowledge G.M. D'Ariano, G. Introzzi, P.
Perinotti, and W.H. Zurek for interesting discussions and/or useful
feedback. In particular, P. Perinotti suggested the appendix.

\appendix\section{APPENDIX A: Hidden variable models}
This appendix is addressed only to more advanced readers. In the
spirit of the original proof of Bell's theorem
\cite{speakable,ghirardi}, one can relax both the ``counterfactual
definiteness'' and the ``Einstein locality'' hypotheses somewhat. In
fact, instead of supposing that there are some pre-existing properties
of the objects (counterfactual definiteness), we can suppose that the
properties are not completely pre-determined, but that a hidden
variable $\lambda$ exists and the properties have a probability
distribution that is a function of $\lambda$. The ``hidden variable
model'' hypothesis is weaker than counterfactual definiteness: if the
properties are pre-existing, then their probability distribution in
$\lambda$ is trivial: there is a value of $\lambda$ that determines
uniquely the property, e.g.~a value $\lambda_0$ such that the
probability $P_i(a=0|A,\lambda_0)=1$ and hence
$P_i(a=1|A,\lambda_0)=0$, namely it is certain that property $A$ for
object $i$ has value $a=0$ for $\lambda=\lambda_0$.  

We can also relax the ``Einstein locality'' hypothesis, by simply
requiring that the probability distributions of measurement outcomes
factorize (``Bell locality'' \cite{speakable,ghirardi,belloc}).  Call
$P(x,x'|X,X',\lambda)$ the probability distribution (due to the hidden
variable model) that the measurement of the property $X$ on the first
object gives result $x$ and the measurement of $X'$ on the second
gives $x'$, where $X,X'=A,B,C$ denote the three two-valued properties
$A$, $B$, and $C$. By definition, ``Bell locality'' is the property
that the probability distributions of the properties of the two
objects factorize, namely
\begin{eqnarray}
P(x,x'|X,X',\lambda)=P_1(x|X,\lambda)P_2(x'|X',\lambda)
\labell{fact}\;,
\end{eqnarray}
the factorization of the probability means that the probability of
seeing some value $x$ of the property $X$ for object 1 is independent
of which property $X'$ one chooses to measure and what result $x'$ one
obtains on object 2 (and viceversa). The ``Bell locality'' condition
\eqref{fact} is implied by (and, hence, it is weaker than) Einstein
locality. In fact, Einstein locality implies that the measurement
outcomes at one system cannot be influenced by the choice of which
property is measured on a second, distant, system. So, the probability
of the outcomes of the first system $P_1$ must be independent of the
choice of the measured property of the second system $X'$, namely
$P_1(x|X,X',\lambda)=P_1(x|X,\lambda)$. The same reasoning applies to
the second system, which leads to condition \eqref{fact}.

Following \cite{ghirardi}, we now show that a Bell-local, hidden
variable model together with the request that the two systems can have
identical properties, implies counterfactual definiteness. This means
that we can replace ``counterfactual definiteness'' with ``hidden
variable model'' in the above proof of Bell theorem, which, with these
relaxed hypothesis states that ``no local hidden variable model can
represent quantum mechanics''.

If two objects have the same property, then $P_{same}(X,X)=1$, namely
the probability that a measurement of the same property $X$ on the two
objects gives opposite results (say, $x=1$ and $x'=0$) is null. In
formulas,
\begin{eqnarray}
  \sum_\lambda P(x=1,x'=0|X,X,\lambda)\;p(\lambda)=0\;,
\labell{pr}\;
\end{eqnarray}
where the $\sum_\lambda$ emphasizes that we are averaging over the
hidden variables (since they are hidden): $p(\lambda)$ is the
probability distribution of the hidden variable $\lambda$ in the
initial (joint) state of the two systems. Note that in Eq.~\eqref{pr}
we are measuring the same property $X$ on both objects but we are
looking for the probability of obtaining opposite results $x'\neq x$.
Using the Bell locality condition \eqref{fact} the probability
factorizes, namely Eq.~\eqref{pr} becomes
\begin{eqnarray}
  \sum_\lambda P_1(x=1|X,\lambda)\;P_2(x'=0|X,\lambda)\;p(\lambda)=0
\labell{pr1}\;.
\end{eqnarray}
Since $P_1$, $P_2$, and $p$ are probabilities, they must be positive.
Consider the values of $\lambda$ for which $p(\lambda)>0$: the above
sum can be null {\em only if} either $P_1$ or $P_2$ is null. Namely if
$P_1(x=1|X,\lambda)=0$ (which implies that $X$ has the predetermined
value $x=0$) or if $P_2(x'=0|X,\lambda)=0$ (which means that $X$ has
predetermined value $x'=1$): we remind that counterfactual
definiteness means that $P_i(x|X,\lambda)$ is either 0 or 1: it is
equal to 0 if the property $X$ of the $i$th object does not have the
value $x$, and it is equal to 1 if it does have the value $x$.  We
have, hence, shown that Eq.~\eqref{pr1} implies counterfactual
definiteness for property $X$: its value is predetermined for one of
the two objects.

Summarizing, if we assume that a Bell-local hidden variable model
admits two objects that have the same values of their properties, then
we can prove counterfactual definiteness. This means that we can relax
the ``counterfactual definiteness'' and ``Einstein locality''
hypotheses in the proof of the Bell theorem, replacing it with the
``existence of a hidden variable model'' and with ``Bell locality''
respectively, so that the Bell theorem takes the meaning that ``no
Bell-local hidden variable model can describe quantum mechanics'' [the
hypothesis that two objects can have the same values for the
properties is implicit in the fact that such objects exist in quantum
mechanics, see Eq.~\eqref{pa}].  Namely, if we want to use a hidden
variable model to describe quantum mechanics (as in the de
Broglie-Bohm interpretation), such model must violate Bell locality.
Otherwise, if we want to maintain Bell locality, we cannot use a hidden
variable model (as in the Copenhagen interpretation).

\section{APPENDIX B: Summary of the hypotheses used and logic
  formalization of Bell's theorem}
We have given two different proofs of the Bell inequality based on
different hypotheses. In this appendix we summarize the logic behind
the Bell inequality proofs.

Hypotheses we used (rigorously defined above):
\begin{itemize}
\item[(A)] ``Counterfactual Definiteness''.
\item[(B)] ``Einstein locality''.
\item[(C)] ``No super-determinism''
\item[(D)] ``Measurement independence''
\item[(A')] ``Hidden variable model'', implied by (A) and by the fact that
  systems with same properties exist (see Appendix A).
\item[(B')] ``Bell locality'', implied by (B) (see Appendix A).
\end{itemize}
In the
main text we have proven (Fig.~\ref{f:p}) the following theorem:\\
\texttt{(A) AND (B) AND (C) AND (D) $\Rightarrow$ Bell inequality
  $\Rightarrow$ NOT QM},\\
where with ``\texttt{NOT QM}'' we mean that quantum mechanics (QM)
violates the Bell inequality and is, hence, incompatible with it.
Using the fact that ``\texttt{X AND Y $\Rightarrow$ NOT Z}'' is
equivalent to ``\texttt{Z $\Rightarrow$ NOT X OR NOT Y}'' ({\em modus
  tollens}), we can state the above theorem equivalently as \texttt{QM
  $\Rightarrow$ NOT (A) OR NOT (B) OR NOT (C) OR NOT (D)}. Since one
typically assumes that both (C) and (D) are true, they can be dropped
and the theorem can be written more compactly as\\
\texttt{QM $\Rightarrow$ NOT (A) OR NOT (B).}\\
Namely, (assuming ``no super-determinism'' and ``measurement
independence'') \texttt{quantum mechanics implies that either
  ``counterfactual definiteness'' or ``Einstein locality'' must be
  dropped}.  This is the most important legacy of Bell.

We have also seen that the hypotheses (A) and (B) can be weakened
somewhat, so that the Bell inequality can also be derived
using only (A') and (B'). Namely, we can prove (see Appendix A):\\
\texttt{(A') AND (B') AND (C) AND (D) $\Rightarrow$ Bell inequality
  $\Rightarrow$ NOT QM}.\\
Namely, (assuming ``no super-determinism'' and ``measurement
independence'') quantum mechanics is incompatible with Bell-local
hidden variable models.


\begin{references}
\bibitem{epr}A. Einstein, B. Podolsky, N. Rosen, {\em Can
    quantum-mechanical description of physical reality be considered
    complete?}, Phys. Rev. {\bf 47}, 777 (1935).
\bibitem{schilpp}A. Einstein, in {\em A. Einstein,
    Philosopher-Scientist}, ed. by P.A. Schilpp, Library of Living
  Philosophers, Evanston (1949), pg. 671.
\bibitem{bell}J. S. Bell, {\em On the Einstein Podolsky Rosen
    Paradox}, Physics {\bf 1}, 195 (1964); Bell J. S., {\em On the
    problem of hidden variables in quantum mechanics}, Rev. Mod. Phys.
  {\bf 38}, 447 (1966).
\bibitem{speakable} J. S. Bell, {\em Speakable and Unspeakable in
    Quantum Mechanics} (Cambridge Univ. Press, Cambridge, 1987).
\bibitem{merminrmp} N.D. Mermin, {\em Hidden variables and the two
    theorems of John Bell}, Rev. Mod. Phys. {\bf 65}, 803 (1993).
\bibitem{werner} ``Bell's telephone'' in R. Werner, {\em Quantum
    Information Theory---an Invitation}, Springer Tracts in Modern
  Physics {\bf 173}, 14, (2001), available from
  http://arxiv.org/pdf/quant-ph/0101061.
\bibitem{peres}A. Peres, {\em Unperformed experiments have no
    results}, Am. J. Phys. {\bf 46}, 745 (1978).
\bibitem{mermin}N.D. Mermin, {\em Bringing home the atomic world:
    Quantum, mysteries for anybody}, Am. J. Phys. {\bf 49}, 940
  (1981).
\bibitem{preskill} J. Preskill, lecture notes at
  http://www.theory.caltech.edu/people/preskill/ph229.
\bibitem{auletta}G. Auletta, {\em Foundations and Interpretation of
    Quantum Mechanics} (World Scientific, Singapore, 2000).
\bibitem{kwiat}P.G. Kwiat, L. Hardy, {\em The mystery of the quantum
    cakes}, Am. J. Phys. {\bf 68}, 33 (2000).
\bibitem{wiseman}Kurt Jacobs and Howard M. Wiseman, {\em An entangled
    web of crime: Bell's theorem as a short story}, Am. J. Phys. {\bf
    73}, 932 (2005).
\bibitem{macdon}A.L. Macdonald, {\em Comment on "Resolution of the
    Einstein-Podolsky-Rosen and Bell Paradoxes}, Phys. Rev. Lett. {\bf
    49}, 1215 (1982).
\bibitem{perbell} A. Peres, {\em Existence of ``Free will'' as a
    problem of Physics,} Found. Phys. {\bf 16}, 573 (1986).
\bibitem{stapp} H.P. Stapp, {\em S-Matrix Interpretation of Quantum
    Theory,} Phys. Rev. D {\bf 3}, 1303 (1971); {\em Bell's theorem
    and world process}, Nuovo Cimento {\bf 29B}, 270 (1975); {\em Are
    superluminal connections necessary?}, Nuovo Cimento {\bf 40B}, 191
  (1977); {\em Locality and reality,} Found. Phys.  {\bf 10}, 767
  (1980); W. De Baere, {\em On Some Consequences of the Breakdown of
    Counterfactual Definiteness in the Quantum World,} Fortschr. Phys.
  {\bf 46}, 843 (1998).
\bibitem{botero}Y. Aharonov, A. Botero, M. Scully, {\em Locality or
    non-locality in quantum mechanics: Hidden variables without
    ``spooky action-at-a-distance''}, Z. Natureforsh. A {\bf 56}, 5
  (2001).
\bibitem{grangier}A. Aspect, P. Grangier, G. Roger, {\em Experimental
    Realization of Einstein-Podolsky-Rosen-Bohm Gedankenexperiment: A
    New Violation of Bell's Inequalities}, Phys. Rev. Lett. {\bf 49},
  91 (1982).
\bibitem{kwiatex}P.G. Kwiat, E. Waks, A.G. White, I.  Appelbaum, P.H.
  Eberhard, {\em Ultrabright source of polarization-entangled
    photons}, Phys. Rev. A {\bf 60}, R773 (1999).
\bibitem{zeil}M. Giustina, A. Mech, S. Ramelow, B.  Wittmann, J.
  Kofler, J. Beyer, A. Lita, B. Calkins, T. Gerrits, S. Woo Nam, R.
  Ursin, A. Zeilinger, {\em Bell violation with entangled photons,
    free of the fair-sampling assumption}, Nature {\bf 497}, 227-230
  (2013).
\bibitem{pearle}P.M. Pearle, {\em Hidden-Variable Example Based upon Data
  Rejection}, Phys. Rev. D {\bf 2}, 1418 (1970).
\bibitem{bellc}J.S. Bell, {\em Atomic-cascade Photons and
    Quantum-mechanical nonlocality}, Comments Atom. Mol. Phys. {\bf
    9}, 121 (1980); reprinted as Chap.~13 of \cite{speakable}.
\bibitem{ghirardi}G. Ghirardi, {\em On a recent proof of nonlocality
    without inequalities}, Found. Phys. {\bf 41}, 1309 (2011), also at
  arXiv:1101.5252.
\bibitem{belloc}T. Norsen, {\em Bell Locality and the Nonlocal
    Character of Nature}, Found. Phys. Lett. {\bf 19}, 633 (2006).
\end{references}
\end{document}